\documentclass[prd,nofootinbib,onecolumn,11pt]{revtex4}
	   \usepackage{amsmath}
	  \usepackage{amssymb,amsthm}
	  \usepackage{graphics}
	  \usepackage{epsfig}
	  \usepackage{feynmp}
	  \usepackage[dvipdfm]{hyperref}
	  \usepackage[dvips]{color}
	  \usepackage{calc}
	  \usepackage{latexsym}
	  \usepackage{epsfig}
	  \usepackage{mathrsfs}
	  \usepackage{default}
	  \usepackage{floatflt}
	  \usepackage[active]{srcltx}

	    \newcommand\nn{\nonumber}
	    \newcommand{\ba}{\begin{eqnarray}}
	    \newcommand{\ea}{\end{eqnarray}}

\begin{document}

	     \title{Radiative corrections to the spectra of  heavy particles  annihilation into leptons: Dark matter implementation.}

             \author{A.~Radovskaya$^{1,2}$}   
	     \email{raan@theor.jinr.ru}  
             \author{E. A.~Kuraev$^1$}
	     \affiliation{$^1$Joint Institute for Nuclear Research, Dubna, Russia\\
                  $^2$Irkutsk State University, Irkutsk, Russia}

	    \date{\today}

	    \begin{abstract}
            We calculate the radiative corrections to the annihilation  
            of the Dark Matter particles into leptons.
	    Lepton masses are taken into account.  
            As the Dark Matter particles  are considered  both Dirac  
	    and Majorana fermions. We  sum up  all the leading logarithms contributions where
            it is possible. We show that in the Dirac fermion case the resulting cross-section
            is consistent with the renormalization group prediction:
	    the cross-section have the Drell-Yan form. Whereas, in the Majorana fermions case
            we have two essentially different limits of the resulting cross-section and quantitively the answer strongly 
             depends on  a spectrum of the masses 
	    entering the problem.
	     \end{abstract}


	    \maketitle

	     \section{Introduction}

	    It is well known fact that  approximately  quarter of the energy density of our universe belongs to
	    the invisible Dark Matter. The Dark Matter particles are supposed to be  neutral heavy particles 
	    so we can not observe  them in the sky directly. One can detect them  through their gravitational influence 
	    on the visible matter. But also we can observe the products of the dark matter particles annihilation 
	    such as gamma rays, positrons, antiprotons, neutrinos, etc. 
            (so called indirect method of the dark matter detection).

	    Possible detection of the Dark Matter particles is one of the most exciting problems of the astroparticle physics 
            \cite{astro}.	
	    A lot of experiments have found new phenomena 
            which might be considered as the traces of the new particles \cite{experiment}.
	    To catch new particles one should  know the annihilation cross-section, 
	    the distribution of the Dark Matter in the Galactic halo and the model of the propagation of resulting particles 
	    from the interaction point to our detectors. 
	    Putting aside astrophysical questions we concentrate on the cross-sections calculation. 

	    In particular, in this work we investigate the  process of the Dark Matter particles annihilation 
            into the lepton pair 
	    with respect to the radiative corrections. We are interested in both the Dirac and Majorana fermions  
	    as the Dark Matter candidates.
	    Such a process with respect to the Majorana particles have been calculated 
	    in the paper by Bergstrom at al.\cite{bergstrom} and interesting results have been obtained. 
	    It was stated that for the Majorana fermion annihilation the contribution 
	    of the next-to-leading order terms can increase the cross-section significantly. 

	    In our calculation we  reproduce result \cite{bergstrom}, 
	    investigate dependence on a lepton mass and show  the renormalization group structure of this cross-sections. 

	    The paper is organized as follows: after the brief review of the Born cross-sections 
            for the both Dirac and Majorana cases in section [\ref{born}] we calculate the soft
            and virtual photon corrections in section [\ref{soft}] and consider the hard photon emission corrections
            in section [\ref{hard}] in details. After that in section [\ref{drell}] we demonstrate  
            the Drell-Yan form of the both cross-sections with the large mass of the intermediate particle.  
            We summarise our results and conclusions in the last section [\ref{res}].

          \section{Born cross sections\label{born}}
	      In our calculations we use interaction Lagrangians where $\Psi$ is the electron  field
	    and $\chi_D$ - the Dirac Dark Matter fermion:
	      \ba
		\mathcal{L}_D^{int}=g_L \bar\Psi\hat R \chi_D \phi_L +g_L^* \bar \chi_D\hat L \Psi \phi_L^* 
		+ g_R \bar\Psi\hat L \chi_D \phi_R +g_R^* \bar \chi_D\hat R \Psi \phi_R^*,\nn
	      \ea 
	    and for Majorana fermion:
	      \ba
		\mathcal{L}_M^{int}= g_L \bar\Psi\hat R \chi_M \phi_L +g_L^* \chi_M^T C\hat L \Psi \phi_L^* 
		+ g_R \bar\Psi\hat L \chi_M \phi_R +g_R^* \chi_M^T C\hat R \Psi \phi_R^*.
		\label{majorana_lagrangian}
	      \ea 
	    Here $\chi_M$ is the Majorana field, C -the charge conjugation matrix,
	    $\hat L=(1-\gamma_5)/2$, $\hat R=(1+\gamma_5)/2$ - the projection operators and 
            $\phi_R$ and $\phi_L$ are two different scalar fields.
	    The Lagrangian (\ref{majorana_lagrangian}) simulates the part of MSSM 
	    \footnote{MSSM - Minimal Supersymmetric Standard Model} Lagrangian \cite{mssm}
	    i.e. interaction of neutralino ($\chi_M$) with electron ($\Psi$ ) and selectron ($\phi_{L,R}$).
	    We suppose, for simplisity, that masses of the scalars $\phi_R$ and 
	    $\phi_L$ (selectrons in MSSM) are identical and equal to $M_s$. 
	    The Dark Matter particle mass M is less or equal to $M_s$.
            
            Further, for simplicity, we put coupling  $g_R$ to zero. The result for the full Lagrangian
            will be shown in section [\ref{res}].

            \begin{figure}[h]
             \centering
              \includegraphics[height=0.14\textheight]{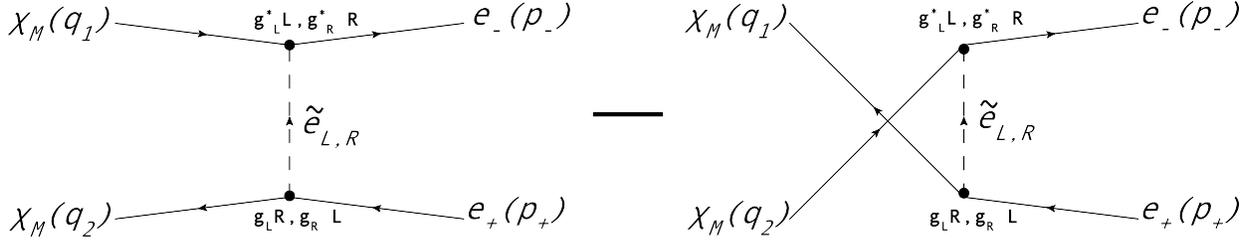}
	       \caption{The annihilation cross-section of the Majorana particles in the Born approximation}
                \label{majorana_born_pict}
            \end{figure}

            Also, in the light of the MSSM, one can call Majorana fermion neutralino, 
            intermediate scalar particles - sleptons.

	    At first, let us consider the neutralino annihilation in the Born approximation (Fig.\ref{majorana_born_pict}):
	    \ba
	    \chi_M(q_1)+\chi_M(q_2)\to e^-(p_-)+e^+(p_+).
	    \ea             
          
		Because of indistinguishability of incoming neutralinos one should  add the cross  graph
	    apart  from the direct one (Fig.\ref{majorana_born_pict}). 
            The "minus" sign  goes from the fermionic nature of the neutralinos. 
	    There are a lot of  papers describing Feynman rules for the Majorana fermions (in different ways),
	    see, for example \cite{CHUNG} or \cite{two}, but one can easy write down the matrix element for this process 
            directly   using 
	    Wick's theorem. Anyway as a result we have:
	    \ba
	    M_M^B=M_{direct}-M_{cross}=
            \frac{|g_L|^2}{t} 
                 \bar u(p_-)\hat R u(q_1)\cdot \bar v(q_2) \hat L v(p_+) 
                           - \frac{|g_L|^2}{\tilde t}\bar u(p_-) \hat R u(q_2)\cdot \bar v(q_1)\hat L v(p_+)                    
	      \ea
	    Where, as usual, $t=q^2-M_s^2$ and $q=q_1-p_-=p_+-q_2$ is the  transfered momentum,  \\
	    correspondingly  $\tilde t=\tilde q^2-M_s^2$, $\tilde q=q_2-p_-=p_+-q_1$ and
	   $q_{1,2}$ are the four-momenta of the initial particles supposed to be
		slow (center of mass is implied):
	    \ba
	    q_1=(E, P, 0,0),\quad q_2=(E,-P,0,0),\quad \frac{P}{E}=v\ll 1,\quad q_1^2=q_2^2=M^2,\quad
	    p_+^2=p_-^2=m^2,   \nn \\
	         M^2 \gg m^2,\quad   t=(p_--q_1)^2-M_s^2=-M^2 a,\quad a=1+\frac{M_s^2}{M^2},
	    \ea
	    where $M$ and $M_s$ are masses of the incoming and intermediate particles respectively.
	    The Moller factor in the cross sections  is assumed to be
	    $I=\sqrt{(q_1q_2)^2-M^4}=2M^2v$ everywhere, (here $v$ is the velocity of the incoming particles
	    in the center-of-mass frame). 

	    We consider the heavy annihilating particles to be almost at rest:
	    \ba
	    q=q_1=q_2=M(1,0,0,0),\quad p_\pm=(E,\pm {\bf p}),\quad  E^2-{\bf p}^2=m^2,\nn \\
	    E=M,\quad t=\tilde t=-M^2a.
	    \ea
             Using the formulae given in the Appendix we obtain:
	    \ba
	    \frac{1}{4}\sum_{spins} |M_{direct}|^2=\frac{1}{4}\sum_{spins} |M_{cross}|^2=\frac{|g_L|^4}{t^2}M^2E^2,\quad
	    \frac{1}{4}\sum_{spins} Re M_{direct}M_{cross}^*=\frac{|g_L|^4}{t^2}M^2(p_+p_-).
	    \ea
	    For the squared total matrix element  we have:
	    \ba
	   \frac{1}{4} \sum_{spins}|M_M^B|^2=\frac{|g_L|^4}{a^2M^2}[2E^2-(p_+p_-)]=\frac{m^2|g_L|^4}{a^2M^2}.
	    \ea

	    The Born cross-section, respectively:  
	      \ba
		\sigma_M^B=\frac{\frac{1}{4}\sum|M_M^B|^2 \mbox{d}\Gamma_2}{4 I}=\frac{|g_L|^4 m^2}{2^6\pi v a^2 M^4}.
	      \ea

	    So, it is the famous result of the cancellation in the Born cross-section 
	    of the annihilating Majorana fermions on the small lepton mass  \cite{gold}. 
	    This cross-section becomes negligible with  lepton as light as the electron. 

\begin{figure}[h]
		\centering
		  \includegraphics[height=0.14\textheight]{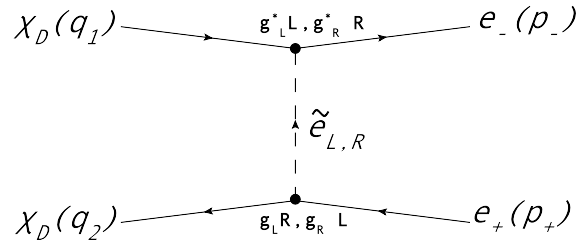}
		  \caption{The annihilation cross-section of the Dirac particles in the Born approximation}
		    \label{dirac_born_pict}
	      \end{figure}

	    The Born cross-section for the annihilation of the Dirac particles is rather easy. One need to calculate 
            only one Feynman diagram (Fig.\ref{dirac_born_pict}).

	      \ba
	      \sigma^{B}_D=\frac{|g_L|^4}{2^6\pi v a^2  M^2}.
	      \ea
	    Here $a=1+(M_s^2/M^2)$, where $M$ and $M_s$ are the masses of the annihilating 
             and intermediate particles respectively. As one can see, in the case of the annihilation 
            of the Dirac particles there is no suppression on a small lepton mass. 

	 \section{Contribution from the emission of the  virtual and soft real photons\label{soft}}

            Now, when  we have leaned the  Born cross sections for the processes, we are interested in 
            the calculation of the perturbative corrections 
            related  with the emission of the hard or soft real and virtual photons. 
            In this section we take into account the last two processes which are universal
            for both  Dirac and Majorana cases.
               \subsection*{the virtual photons and the counterterms}
\begin{figure}[h]
		\centering
		  \includegraphics[height=0.11\textheight]{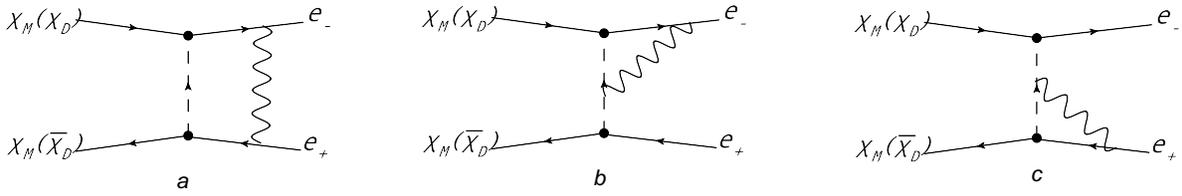}
		  \caption{The virtual photon correction to the annihilation cross-section}
		    \label{rad_cor_pict}
		\end{figure}

	    Among the one-loop Feynman diagrams the main contribution (containing the so-called
	    "Large Logarithm" terms) arises from the QED type ones. As soon as we 
            provide the calculations
	    in frames of unrenormalized theory we must include the ultraviolet cut-off parameter
	    $\Lambda$, perform the loop integration and take into account
	    the electron and positron wave function counterterms \cite{AKHBER}:
		\ba
		&&|M^B|^2 \rightarrow |M^B|^2Z_{e_-}Z_{e_+}, \\
		Z_{e_-}&=&Z_{e^+}=1-\frac{\alpha}{2\pi}\left[\frac{1}{2}\ln\frac{\Lambda^2}{M^2}-\ln\frac{m^2}{\lambda^2}
		+\frac{1}{2}L+\frac{9}{4}\right]=
		S_{EW}\left[1-\frac{\alpha}{2\pi}(\frac{1}{2}L-\ln\frac{m^2}{\lambda^2}+\frac{9}{4})\right], \nn \\
		S_{EW}&=&1-\frac{\alpha}{4\pi}\ln\frac{\Lambda^2}{4M^2},\nn \\
		L&=&\ln\frac{4M^2}{m^2},
		\ea
	    here $\lambda$ is "photon mass" - small auxiliary parameter which will be dropped away in the final
	    step of the calculations. $L$ is so called "large logarithm". The factor $S_{EW}$ will be absorbed
	    by the renormalization of the coupling constant $g$:
	    $$ g^{bare}S_{EW}=g. $$
	    It results into the counterterms contribution to the cross section:      
		\ba
		\frac{\sigma^C}{\sigma^B}=-\frac{\alpha}{\pi}\left[\frac{1}{2}L-
		\ln\frac{m^2}{\lambda^2}+\frac{9}{4}\right].
		\ea
            Notice that the contributions arising from the vertex type Feynman diagrams which are 
            shown in Fig.\ref{rad_cor_pict}b and 
            Fig.\ref{rad_cor_pict}c 
            cancel each other due to the different sign of the electron and positron charges.

	    In the calculations of the contribution from the box type  Feynman diagram (Fig.\ref{rad_cor_pict}a) one
            can assume
	    that loop momentum is small compared to the mass of the intermediate particle $ |k^2|<<M_s^2$:
	    \ba
	    (q-k)^2-M_s^2\approx t.
	    \ea

	    In such a way we
	    will have the amplitude corresponding to the triangular  Feynman diagram. The part of its
	    contribution which contains the logarithm of the ultra-violet cut-off parameter $\ln(\Lambda^2/M^2)$ 
	    will be absorbed as well by the coupling constant renormalization.

	    Interference of this part of 1-loop corrected amplitude with Born one contains:
	    \ba
	    2M^B*M^{box}= -\frac{1}{t^2}\frac{\alpha}{2\pi}\int\limits_0^1 dx\int\limits_0^1\frac{y dy}{d}S_1*S_2,
	    \ea
	    with $d=y^2p_x^2+(1-y)\lambda^2,\ p_x=xp_--(1-x)p_+$ and
	    \ba
	    S_1*S_2={\rm Tr}\left[ \hat{p}_-\gamma_\mu(\hat{p}_--y\hat{p}_x)\hat L(\hat{Q}_1+M)\hat R \right]\times\nn \\
	   {\rm Tr}\left[(\hat{Q}_2-M)\hat L(-\hat{p}_+-y\hat{p}_x)\gamma_\mu\hat{p}_+\hat R\right]. \nn
	    \ea
	    Integration over the  Feynman parameter $y$ can be performed using:
	    \ba
	    \int\limits_0^1\frac{y dy}{d}=\frac{1}{2p_x^2}\ln\frac{p_x^2}{\lambda^2};
	    \qquad
	    \int\limits_0^1\frac{y^2dy}{d}=\frac{1}{p_x^2}.
	    \ea

	    Here $p_x^2=m^2-sx(1-x)-i0,\ s=(p_++p_-)^2=4M^2$ and
	    \ba
	    Re\int\limits_0^1\frac{1}{p_x^2}=-\frac{2}{s}L, \quad
	    Re\int\limits_0^1\frac{1}{p_x^2}\ln\frac{p_x^2}{m^2}=-\frac{1}{s}[L^2-8\xi_2],
	    \qquad
	    \xi_2=\frac{\pi^2}{6},
	    \ea
	    we obtain:
	    \ba
	    \sigma^v=\sigma^B\left(-\frac{\alpha}{\pi}\right)
	    \left[\frac{1}{2}(L^2-8\xi_2)+L\ln\frac{m^2}{\lambda^2}-2L\right].
	    \ea
                    \subsection*{the soft photon emission}
	    Contribution from the soft photon emission is standard:
	    $$
	    \frac{\sigma^{soft}}{\sigma^B}=\left.-\frac{\alpha}{4\pi}\int\frac{d^3k}{\omega}\left(\frac{p_-}{p_-k}-
	    \frac{p_+}{p_+k}\right)^2\right|_{\omega<\Delta E}.
	    $$
	    After integation one can obtain:
	    \ba
	    \frac{\sigma^{soft}}{\sigma^B}=\frac{\alpha}{\pi}\left[2(L-1)\ln\Delta +(L-1)\ln\frac{m^2}{\lambda^2}+
	    \frac{1}{2}L^2-2\xi_2\right],\ \Delta=\frac{\Delta E}{M}<<1.
	    \ea

	    The sum of the virtual and soft photons emission contributions is free from the infrared divergence
	    (remains finite in limit $\lambda\to 0$), as soon as it does not contain terms with $L^2$:
	    \ba
	    \frac{\sigma^C+\sigma^v+\sigma^{soft}}{\sigma^B}=\frac{\alpha}{\pi}\left[(L-1)\left(2\ln\Delta +\frac{3}{2}\right)+
	    2\xi_2-\frac{3}{4}\right].
	    \ea
	    Including the Born cross section the result can be written in form:
	    \ba
	    \frac{\sigma^B+\sigma^C+\sigma^v+\sigma^{soft}}{\sigma^B}=\frac{\sigma^{BSV}}{\sigma^B}=
	    (1+\beta P_\Delta)^2(1+\frac{\alpha}{\pi}K_{vs}), \qquad K_{vs}=2\xi_2-\frac{3}{4},
	    \ea
	    with $P_\Delta=2\ln\Delta+3/2$ is the so called delta-part of the evolution
	    equation kernel $P(x)$ and
	    \ba
	    \beta=\frac{\alpha}{2\pi}(L-1).
	    \ea

	  \section{The hard photon emission\label{hard}}

	    In this section we consider annihilation cross-sections with respect to the hard photon emission, 
            i.e. processes:
           	\ba
		&&\chi_M(q_1)+\chi_M(q_2) \to e^+(p_+)+e^-(p_-)+\gamma(k) \nn \\
    \mbox{and}\ &&\chi_D(q_1)+\bar\chi_D(q_2) \to e^+(p_+)+e^-(p_-)+\gamma(k), \\
                &&\qquad p_\pm^2=m^2, \qquad k^2=0.\nn
		\ea
	    The cross-section for a  2$\to$3 process have a form of:
		\ba
		d\sigma=\frac{\frac{1}{4}\sum |M^{hard}|^2 d\Gamma_3}{4I},
		\ea
	    where the phase volume is:
	    $$ d\Gamma_3=\frac{(2\pi)^4}{(2\pi)^9}\frac{d^3p_+d^3p_-d^3k}{2E_+2E_-2\omega}\delta^4(q_1+q_2-p_+-p_--k),$$
	    can be written in the form:
	    \ba
	    d\Gamma_3=\frac{M^2}{32\pi^3}dx_- dx_+ dx \delta(2-x-x_--x_+),
	    \ea
	    with
	    \ba
	    x_\pm=E_\pm/M, \ x=\omega/M,\ 1-x<x_\pm<1-\delta,
	    \ 
	    \delta=\frac{m^2(1-x)}{4M^2 x},\ 
	    1<x_++x_-<2-\Delta
	    \ea
	  Here  $x>\Delta$ and $\Delta E=M\Delta$ is the minimal value of the photon energy, where the
         photon can be considered as a hard one.

          \subsection*{Majorana case}
             Let us now calculate hard photon emission for the Majorana cross-section.
            \begin{figure}[h]
	    \includegraphics[width=\textwidth]{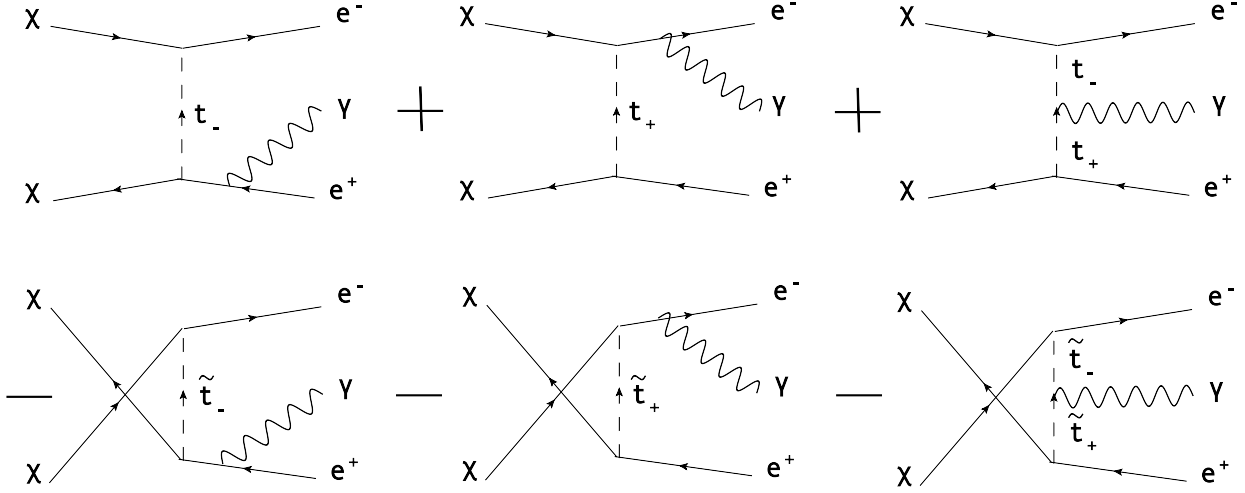}
	    \caption{The hard photon emission for the Majorana cross-section}
	      \label{majorana_hard_rad_pict}
	  \end{figure}
	     As well as for the Majorana Born cross-section, here we have "doubling" of the graphs
        in the matrix element (Fig.\ref{majorana_hard_rad_pict}). So, the appropriate matrix element contains
           6 gauge invariant Feynman amplitudes:
		\ba
		M_M^{hard}&=&-ie (M_{direct}^{hard}-M_{cross}^{hard}),\nn \\
		M_{direct}^{hard}&=&|g_L|^2\left(
		\frac{1}{t_-\chi_+}\bar u(p_-)\hat Lu(q_1)\cdot \bar v(q_2)\hat R(-\hat p_+-\hat k+m)\hat \varepsilon v(p_+)
               +\right.\nn \\
	      &+&\frac{1}{t_+ \chi_-}\bar u(p_-)\hat \varepsilon(\hat p_-+\hat k+m)\hat Lu(q_1) \bar v(q_2)\hat R v(p_+)+ \nn \\
	&+&\left.\frac{(\varepsilon,q_++q_-) }{t_+ t_-}\bar u(p_-)\hat Lu(q_1) \bar v(q_2)\hat R v(p_+)\right),
	      \nn\\
	      M_{cross}^{hard}&=&M_{direct}^{hard}(t_{\pm}\rightarrow\tilde t_{\pm}, q_1\leftrightarrow q_2).
              \label{M_direct}
	      \ea
               Here and below $t_{\pm}=q_{\pm}^2-\tilde m^2$, where $q_+=p_+-q_2,\ q_-=q_1-p_-\ $,
         $\tilde t_{\pm}=\tilde q_{\pm}^2-\tilde m^2$,\\ where $\tilde q_+=p_+-q_1,\ \tilde q_-=q_2-p_-$.
  
	 Now we need to square this matrix element. It consists of three essentially different parts: \\ 
         $A_{pole}$ - the part that have poles when $x_+$ or $x_-$ are equal to 1,
         $A_0$ - part remaining in the  zero lepton mass limit and $A_{other}$ - 
         other terms which are insignifficant because of
         their smallness as we will see later. 
          \ba
            \frac{1}{4}\sum_{spins}|M_M^{hard}|^2&=&A_{pole} +A_0 + A_{other},\\
             A_{pole}&=&\frac{e^2|g_L|^4 m^2}{M^4}\left(
                      \frac{a+2x_-+x_-(a+2x_+)}{2 (1-x_+)\tau_+^2\tau_-}
                     +\frac{a+2x_++x_+(a+2x_-)}{2 (1-x_-)\tau_+\tau_-^2}\right.\nn \\
                     &+&\left.\frac{1}{(1-x_+)(1-x_-)\tau_+\tau_-}-\frac{m^2}{4(1-x_+)^2\tau_+M^2}-
                     \frac{m^2}{4(1-x_-)^2\tau_-M^2}\right).
          \ea
          Where $\tau_{\pm}=2(1-x_{\pm})-a$ goes from propogators of the intermediate particle $t_{\pm}=M^2 \tau_{\pm}$.
	    \ba
	      A_0=\frac{4 e^2 |g_L|^4(1-x)}{M^2 \tau_+^2\tau_-^2}\left[ (1-x_+)^2+(1-x_-)^2\right].
	    \ea
	  As one can see $A_0$ does not contain lepton mass, so this term  can not be proportional 
	  to the Born cross-section for the Majorana particles. The term remains finite in the zero lepton mass limit.
	 
	    \ba
	  A_{other}=\frac{e^2|g_L|^4 m^2}{2 \tau_+^2\tau_-^2M^4}((\tau_+-\tau_-)^2+\tau_+^2+\tau_-^2 -6(\tau_+ +\tau_-)+\nn \\
	  24(x_++x_-)-16x_+x_--24 )+O(\frac{m^4}{M^6}).
           \ea
          In order to explicitly separate pole structure of our result, i.e. future "Large Logarithms",
           we rewrite $A_{pole}+A_{other}=A_{sing}+\tilde\Phi_M$ with help of the followng identities:
            \ba
            \frac{f(x)}{x-x_0}&=&\frac{f(x_0)}{x-x_0}+\frac{f(x)-f(x_0)}{x-x_0} ,\nn \\
            \frac{f(x)}{(x-x_0)^2}&=&\frac{f(x_0)}{(x-x_0)^2}+\frac{f(x)-f(x_0)-f'(x_0)(x-x_0)}{(x-x_0)^2}+O(1),
            \ea  
            \ba     
              \frac{1}{(1-x_+)(1-x_-)}=\frac{1}{x}\left(\frac{1}{1-x_+}+\frac{1}{1-x_-}\right), 
              \ \mbox{remind that}\ 2=x+ x_+ +x_-.
            \ea
           Thus we arrive to the expression:
          \ba
             \frac{1}{4}\sum_{spins}|M_M^{hard}|^2&=&A_{sing} +A_0 + \tilde\Phi_M,\\
            A_{sing}&=&\frac{e^2|g_L|^4 m^2}{2 a^2 M^4}\left(\frac{2-2x}{x}-\frac{ax^2}{x(2x-a)} \right)
            \left[\frac{1}{1-x_+}+\frac{1}{1-x_-}\right]-\nn \\
            &-&\frac{e^2|g_L|^4 m^4}{4a^2M^6}\left[\frac{1}{(1-x_+)^2}+\frac{1}{(1-x_-)^2}\right].\nn\\
          \ea
         Here $\tilde\Phi_M$ is the part of so-called K-factor for Majorana case.
         One can find it the Appendix. 

	    Notice that in the limit of the large intermediate particle mass ($a\gg 2$) one has for singular term:
          \ba
            A_{sing}\to A_{\gamma}=\frac{e^2|g_L|^4 m^2}{2 a^2 M^4}\left(\frac{1+(1-x)^2}{x} \right)
            \left[\frac{1}{1-x_+}+\frac{1}{1-x_-}\right]-\nn \\
            -\frac{e^2|g_L|^4 m^4}{4a^2M^6}\left[\frac{1}{(1-x_+)^2}+\frac{1}{(1-x_-)^2}\right].
          \ea
        So we have two essential limits:
	    \begin{equation}
	     \frac{1}{4}\sum_{spins}|M_M^{hard}|^2 =\left\{
               \begin{array}{lr}
               A_0,& \ \mu^2=\frac{m^2}{M^2}\ll 1 \\
              A_M^{\gamma}+\tilde\Phi_M,\ & a\gg 2,
                \end{array}
             \right.
	    \end{equation}
          or in terms of the differential cross section:
             \begin{equation}
	     \frac{\mbox{d}^2\sigma^{hard}_M}{\mbox{d}x_+\mbox{d}x_-}=\left\{
               \begin{array}{lr}
               F_0,& \ \mu^2\ll 1 \\
              F_M^{\gamma},\ & a\gg 2,
                \end{array}
             \right.
              \label{sigma_hard_M}
       	    \end{equation}
          with 
         \ba
            F_0=\frac{\alpha|g_L|^4(1-x)}{2^4 \pi^2 v\tau_+^2\tau_-^2 M^2 }\left[ (1-x_+)^2+(1-x_-)^2\right],
         \label{f0}
         \ea
         \ba
            F_M^{\gamma}=\sigma_M^B\frac{\alpha}{\pi}\left\{\left(\frac{1+(1-x)^2}{2x} \right)
            \left[\frac{1}{1-x_+}+\frac{1}{1-x_-}\right]
            -\frac{\mu^2}{4}\left[\frac{1}{(1-x_+)^2}+\frac{1}{(1-x_-)^2}\right]+\Phi_M\right\}.
         \ea
          Here new notations were introduced:
          \ba 
          \Phi_M=\frac{M^2 a^2}{\mu^2}\tilde \Phi_M,  \qquad  \mu=\frac{m}{M}.
          \ea
          
          Now we are interested in the case of the zero lepton masses. Imagine that we put m = 0 from the beginning. 
          Than Born cross-section is equal to zero ($\sigma_M^B=0$). There is no emission from $e^+$ and $e^-$ legs, 
          only from intermediate scalar. Soft real and virtual photon contributions are absent.
          Then we obtain for double differential cross-section:  
          \ba
           \frac{\mbox{d}^2\sigma_0}{\mbox{d}x_+\mbox{d}x_-}=\frac{\alpha|g_L|^4(1-x)}{2^4 \pi^2 v\tau_+^2\tau_-^2 M^2 }
          \left[ (1-x_+)^2+(1-x_-)^2\right]
          \ea 
          Integrating over $x_-$ one can find:
           \ba
             \frac{\mbox{d}\sigma_0}{\mbox{d}x_+}=\frac{\alpha |g_L|^4}{2^8\pi^2 v a \tau_+^2 M^2}
          \left[2x_+(-3 a^2+ax_+-4(1-x_+)^2)\right.\nn \\
           -\left. a(-3 a^2+4ax_+-4(1-x_+)^2)
          \ln\left(\frac{a}{a-2x_+}\right) \right].
         \label{dsM0_dxplus}
           \ea      
          And performing last integration we receive:
          \ba 
            \sigma_0=\frac{\alpha |g_L|^4}{2^7\pi^2 v a  M^2}\left( 4a-1 
            -\frac{a((11-4a)a-6)}{2a-2}\ln\left[\frac{a-2}{a}\right]\right.\nn \\
            +\left.a^2\left(\frac{\pi^2}{6}-\ln^2\left[\frac{a}{2a-2}\right] 
          -2\mathcal{L}i_2 \left[\frac{a}{2a-2}\right] \right)
            \right).
          \ea 
         Here $\mathcal{L}i_2 =-\int\limits_0^1\frac{dt}{t}\ln(1-t)$ is Euler dilogarithm.
         This result reproduces the one given in the paper by Bergstrom at al.\cite{bergstrom}.

	    \subsection*{Dirac case}
\begin{figure}[h]
	    \includegraphics[width=\textwidth]{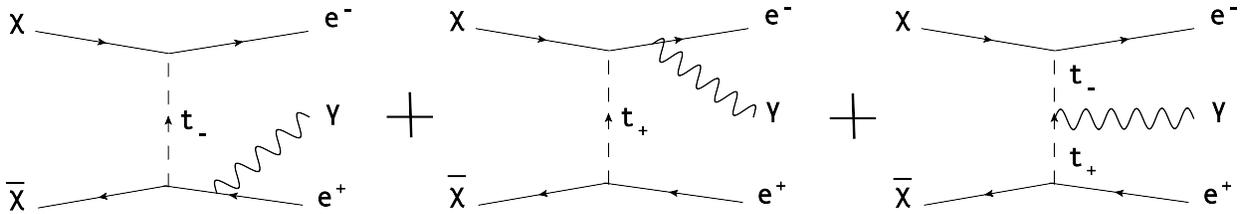}
	    \caption{The hard photon emission for the Dirac cross-section.}
	      \label{dirac_hard_rad_pict}
	  \end{figure}

         Matrix element of the hard photon emission in the case of the Dirac annihilating fermions
          contains only three Feynman diagrams (Fig.\ref{dirac_hard_rad_pict}). It is equal to $M_{direct}^{hard}$
         written down in (\ref{M_direct}).
           Taking the square of the appropriate matrix element and separating singular terms as in previous case we obtain,
          	 in total agreement with the general theory \cite{KF85}, 
	   the following expression:
	    \ba
	     \frac{\mbox{d}^2\sigma^{hard}_D}{\mbox{d}x_+\mbox{d}x_-} &=&\sigma_B^D\frac{\alpha}{\pi} \left\{
                    	   \frac{1+(1-x)^2}{2x}\left(\frac{1}{1-x_+}+\frac{1}{1-x_-}\right)\right.\nn \\
                          &-&\left.\frac{\mu^2}{4}\left(\frac{1}{(1-x_+)^2}+\frac{1}{(1-x_-)^2}\right)+\Phi_D(x_+,x_-)\right\}.
            \label{sigma_hard_D}
	    \ea
          Again,one can find expression for $\Phi_D$  in Appendix. 
          As one can see the structure of the differential cross-section  in Dirac case (\ref{sigma_hard_D}) is almost the same 
          as one in the Majorana case  (\ref{sigma_hard_M})(in the limit a$\gg$2) up to appropriate Born cross-sections. 
           Introduce value
              \ba
	     \frac{\mbox{d}^2\sigma^{hard}_i}{\mbox{d}x_+\mbox{d}x_-} &=&\sigma_i^B\frac{\alpha}{\pi} \left\{
                    	   \frac{1+(1-x)^2}{2x}\left(\frac{1}{1-x_+}+\frac{1}{1-x_-}\right)\right.\nn \\
                          &-&\left.\frac{\mu^2}{4}\left(\frac{1}{(1-x_+)^2}+\frac{1}{(1-x_-)^2}\right)+
               \Phi_i(x_+,x_-)\right\},
            \label{ds_dxdx}
	    \ea
            where i=D,M is specified Dirac and Majorana cases for the appropriate values.
             In such a way we can describe both Dirac and Majorana cases simultaneously.

	    Further integration is straightforward:
	    \ba
	    \int\limits_{1-x_\mp}^{1-\delta}dx_\pm
	    \left[\frac{\mu^2}{(1-x_\pm)^2},\frac{1}{1-x_\pm}\right]=
	    \left[\frac{4x_\mp}{1-x_\mp},L+\ln\frac{x_\mp^2}{1-x_\mp}\right],\\ \quad
            L=\frac{4}{\mu^2},   \quad
	    \delta=\frac{\mu^2 x}{4 x_\mp},\ 2=x+x_++x_-,\ \mu=\frac{m}{M}\nn
	    \ea
            Let us say a few words about the upper limit of the integration. Consider the scalar product 
            of the two four-vectors, for instanse, $(p_-,k)$. On the one hand:
                \ba
            (p_-,k)= E_-\omega - |{\bf p_-}||{\bf k}|\cos \theta_-\approx M^2 x_- x (1-v_-\cos \theta_-),\nn\\
            v_-=\frac{|\bf p_-|}{E_-}=\frac{\sqrt{E_-^2-m^2}}{E_-}\approx 1-\frac{m^2}{2E_-^2}=1-\frac{\mu^2}{2x_-^2}.
                \ea 
            On the other hand:
                \ba
            (p_-,k)=\frac{1}{2}[(p_-+k)^2-m^2]=\frac{1}{2}[(Q_1+Q_2-p_+)^2-m^2]\approx 2 M^2 (1-x_+).
                \ea
            Thus, we have:
                \ba
            x_- x (1-v_-\cos \theta_-)=2 (1-x_+),
                \ea
            and from the constraint $\cos \theta_-<1$ we obtain
                \ba
            x_+<1-\delta, \qquad \delta=\frac{\mu^2 x}{4 x_-}.
                \ea
           Because in the upper limit of the integration $x_+\approx 1$ than $\delta \approx\frac{\mu^2(1-x_-)}{4x_-^2}$.
           Similarly, one can obtain constraint on $x_-$.

	    As a result of the integration we have
            double differential distribution:
	    \ba
	    \frac{d^2\sigma_i^{hard}}{\sigma_B d x_+ d x_-}=\frac{\alpha}{\pi}
	    \left\{\left[\frac{1}{2}\delta(1-x_+)P_\theta(x_-)+
	    \frac{1}{2}\delta(1-x_-)P_\theta(x_+)\right](L-1)+K^h_i(x_+,x_-)\right\},  
            \label{ddd}       
	    \ea
	    with
	    \ba
	    K_i^h(x_+,x_-)=\frac{1}{2}[\delta(1-x_+)(1-x_-+\frac{1+x_-^2}{1-x_-}\ln \frac{x_-^2}{1-x_-})+\\
	    \delta(1-x_-)(1-x_++\frac{1+x_+^2}{1-x_+}\ln \frac{x_+^2}{1-x_+})]+\Phi_i(x_+,x_-)
	    \ea
	    and
	    \ba
	    P_\theta(z)=\frac{1+z^2}{1-z}\theta(1-z-\Delta),\ \Delta=\frac{\Delta E}{M}.
	    \ea
           
	    We note that $P_\theta(z)$ is the so called $\theta$-part  of the
	    evolution equation kernel  for the non-singlet of twist-2 operators matrix elements \cite{KF85}.


            Let us remark here that singularity on lepton mass terms (so-called Large Logarithm
            which can give essential contribution ) arise from such  kinematical region where photon is emitted in the 
           direction close to the charge lepton propogation one.

	    \section{Drell-Yan form of the spectrum\label{drell}}

	    Taking into account the contribution from the emission of the hard photon with  $x>\Delta E/M$
	    we obtain \cite{KF85}:
	    \ba
	    &&\frac{d\sigma_i}{\sigma_i^B}=\frac{d\sigma^{BSV}+d\sigma_i^{hard}}{\sigma_i^B}= \nn \\
	    &=&D(x_-,\beta)D(x_+,\beta)(1+\frac{\alpha}{\pi}K_i(x_+,x_-))dx_+dx_-,\\
            && \beta=\frac{\alpha}{2\pi}(L-1),\ L=\ln\frac{4}{\mu^2},
	    \ea
	    with the Structure function $D(x,\beta))$:
	    \ba
	    D(x,\beta)=\delta(1-x)+\beta P^{(1)}(x)+\frac{1}{2!}(\beta)^2P^{(2)}(x)+... ,
	    \ea
	    and the kernel of evolution equation of twist-2 operators is:
	    \ba
	    P^{(1)}(x)=P(x)=\lim_{\Delta\to 0}[P_\Delta\delta(1-x)+P_\theta(x)\theta(1-x-\Delta)],
	    \ea
	    with
	    \ba
	    P^{(n)}(x)=\int\limits_x^1\frac{dy}{y}P^{(n-1)}(y)P(\frac{x}{y}), n=2,3,... .
	    \ea
	    Kernel $P(x)$ have a property $\int\limits_0^1P(x)dx=0$, which results in
	    \ba
	    \int\limits_0^1 dx D(x,\beta)=1.
	    \ea
	    Using previous relation the inclusive distribution on the positron energy fraction can be
	    obtained as:
	    \ba
	    \frac{d\sigma_i}{\sigma_i^B dx_+}=D(x_+,\beta)(1+\frac{\alpha}{\pi}K_i(x_+)).
            \label{ds_dxplus}
	    \ea
	   
	    Photon energy distribution is (from eq.(\ref{ds_dxdx}))
	    \ba
	    \frac{d\sigma_i}{\sigma_i^B dx}=\int\limits_0^1 dx_+\int\limits_0^1 dx_-\frac{d^2\sigma_i^{hard}}{\sigma_i^B dx_+dx_-}
	    (1+\frac{\alpha}{\pi}K^\gamma_i(x))\delta(2-x-x_+-x_-)
            \label{phod}
             \ea
             The explicit forms of $K_i(x_+)$ and $K^\gamma_i(x)$ factors are:
	    \ba
	    K_i(x_+)&=&\frac{\pi^2}{3}-\frac{3}{4}+\frac{1}{2}\left[1-x_++\frac{1+x_+^2}{1-x_+}\ln \frac{x_+^2}{1-x_+}\right]+ \\
	    &+&\int\limits_{1-x_+}^1\Phi_i(x_+,x_-)dx_-,\ 0<x_+<1, \label{DefK}\nn\\
	    \ea
	    \ba
	    K^\gamma_i(x)&=&\frac{\pi^2}{3}-\frac{3}{4}+x+\frac{1+(1-x)^2}{x}\ln(1-x)+
	    \int\limits_{1-x}^1\Phi_i(2-x-x_-,x_-)dx_-. \label{DefKGamma}
	    \ea
	    %
	    %

	    For practical calculations the smoothed form of the Structure Function can be useful \cite{KF85}:
	    \ba
	    D(x,\beta)=D^{NS}(x,\beta)=2\beta (1-x)^{2\beta-1}[1+\frac{3}{2}\beta]-\beta(1+x)+
	    O(\beta^2).
	    \ea
	    In general the process with two and more electron-positron pairs can be taken into account.
	    It results in the replacement $D^{NS}\to D^{NS}+D^{e^+e^-}$ (see \cite{KF85}).

            \section{results and conclusions\label{res}} 
          In this paper we have considered the Dark Matter particles annihilation into leptons with respect of 
          QED radiative corrections. We have taken into account nonzero leptons masses.  
          In the case of the Dirac incoming fermions  the obtained cross-section is in total agreement with the
          Drell-Yan form. Whereas in the case of the Majorana fermions  the form of the answer 
          strongly depends on the mass spectrum
          of this problem (lepton mass and mass of the intermediate particle).
          In details the results of our calculation are presented below: \\
          {\bf1}) Distribution of the differencial cross-section on the positron energy fraction 
          for Dirac case is (eq.(\ref{ds_dxplus})):
               \ba
	    \frac{d\sigma_D}{dx_+}=\sigma_D^B D(x_+,\beta)(1+\frac{\alpha}{\pi}K_D(x_+))
                \ea
          and for Majorana case is (eqs.(\ref{dsM0_dxplus}) and (\ref{ds_dxplus})):
               \begin{equation}
	     \frac{\mbox{d}\sigma_M}{\mbox{d}x_+}=\left\{
               \begin{array}{lr}
                \frac{\alpha |g_L|^4}{2^8\pi^2 v a \tau_+^2 M^2}
                            \Sigma_M                           ,& \ \mu^2\ll 1 \\
             \sigma_M^B D(x_+,\beta)(1+\frac{\alpha}{\pi}K_M(x_+)) ,\ & a\gg 2,
                \end{array}
             \right.
              \label{dsigma_M}
       	    \end{equation}

       where 
              \ba
                  \Sigma_M=2x_+(-3 a^2+ax_+-4(1-x_+)^2)
           - a(-3 a^2+4ax_+-4(1-x_+)^2)
          \ln\left(\frac{a}{a-2x_+}\right).
              \ea
          Expression for $d\sigma_i$ with the Structure Function $D(x,\beta)$ means that
          effectively we have summarized leading logarithms in all orders of the perturbation series or, physically,
          we have taken into account all the photons emmitted from legs (1 photon + 2 photons + 3 photons + ...)
          \cite{KF85}.

         \begin{figure}[h]
        a\includegraphics[width=3in]{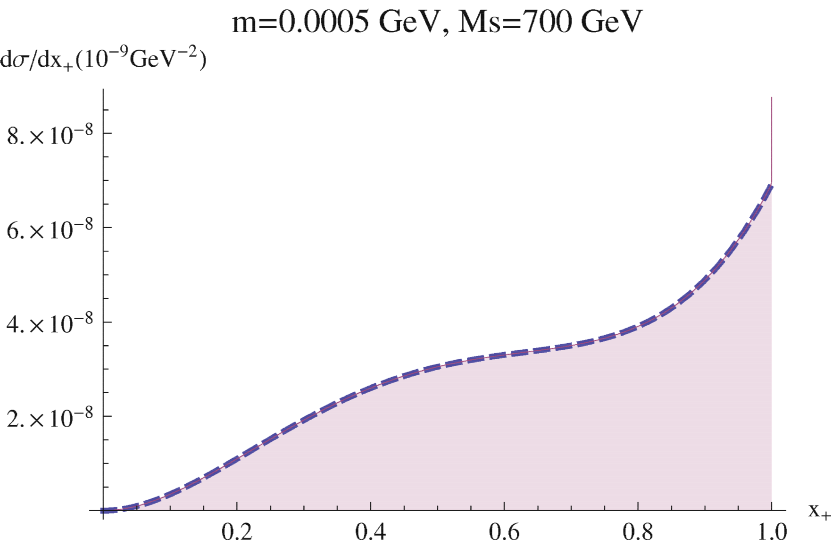}
\hspace{0.5cm}
         b\includegraphics[width=3in]{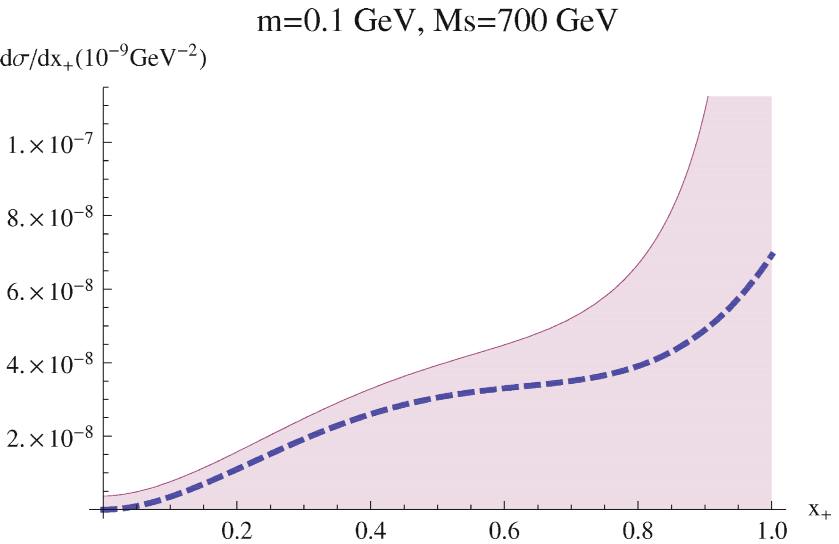}
	 c\includegraphics[width=3in]{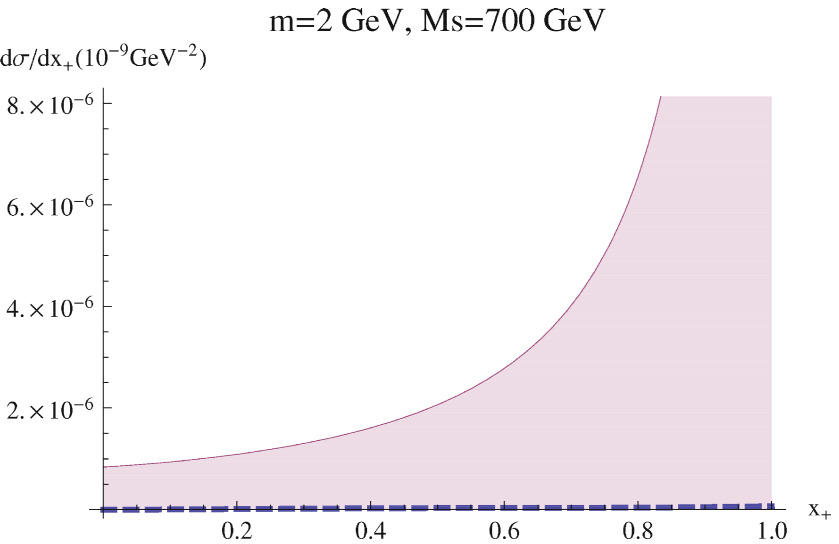}
\hspace{0.5cm}
         d\includegraphics[width=3in]{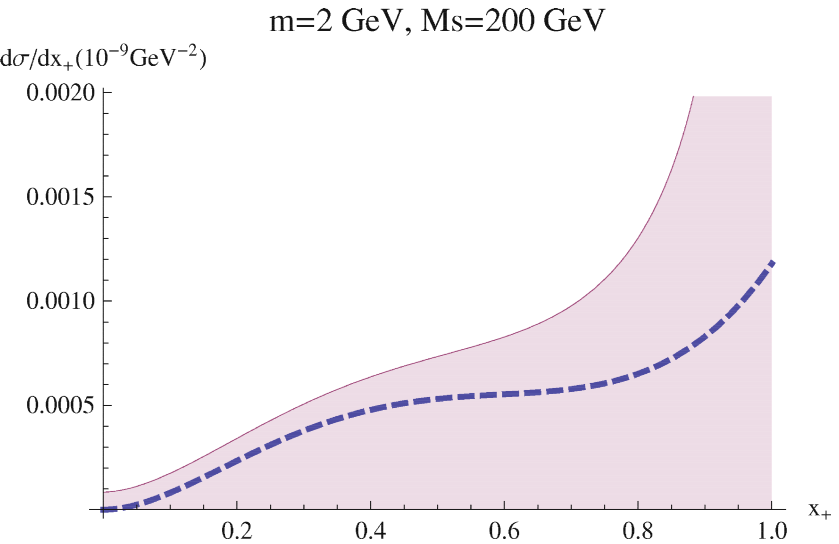}
          e\includegraphics[width=3in]{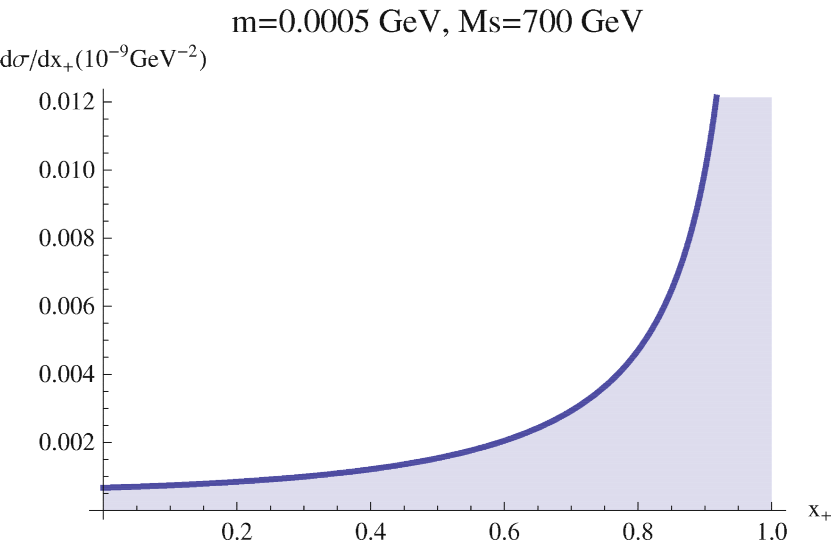}
\hspace{0.5cm}
          f\includegraphics[width=3in]{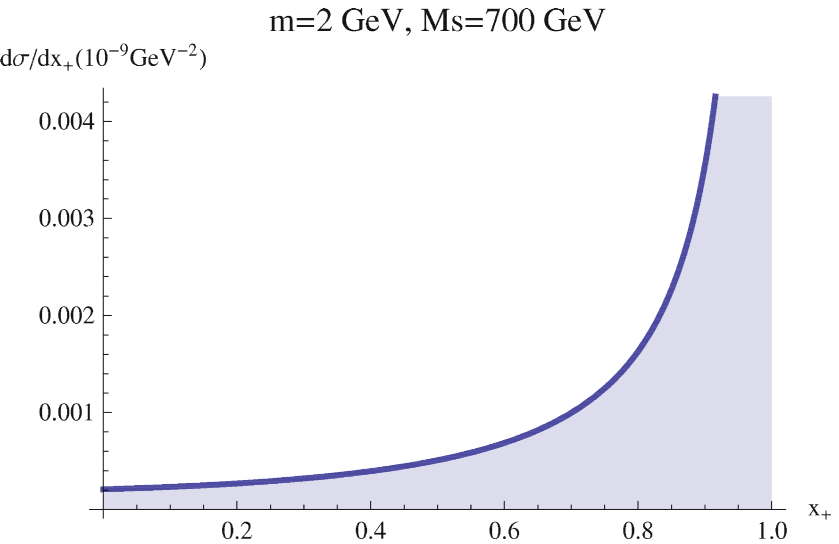}
         \caption{ The distribution of the differencial cross section on the positron energy fraction.
	    \label{s}}
         \end{figure} 
          In the case of the interacting Majorana particles we have two essential limits for the differential cross-section 
          (\ref{dsigma_M}). We have shown the cross-section dependence on masses in Fig.\ref{s}(a-d). Here 
          the filled region denotes the  sum of the cross-sections in both limits $\frac{\mbox{d}\sigma_M}{\mbox{d}x_+}(\mu^2\ll 1)
          + \frac{\mbox{d}\sigma_M}{\mbox{d}x_+}(a\gg 2)$ and dashed line is the contribution of the
          $\frac{\mbox{d}\sigma_M}{\mbox{d}x_+}(\mu^2\ll 1)$ only.
           In other words we have shown the  contribution of the term which is not proportional to the Born cross-section 
         (finite in the zero lepton mass limit) or deviation from the Drell-Yan form of the spectrum. From the mathematical
         point of view we have a seesaw between $m^2 \ln m^2$  and $1/a^2$ terms. 
          As one can see: \\
           -- In the case of the  electron mass  (Fig.\ref{s}a) the Born cross-section becomes negligible and 
          the first corrections exceeds the Born level of the perturbation series. The effect is going to be maximal
          when the intermediate particle is light ($a\sim$2). This effect are interesting in the light of the anomalous 
          positron flow in the Cosmic Rays \cite{bergstrom}.\\
           -- If one takes muon as a lepton and the intermediate particle mass still large enough (Fig.\ref{s}b) than 
           both limits are significant and the Drell-Yan form of the spectrum is still broken.\\
           ---However, when one takes tau-lepton (Fig.\ref{s}c) than still the small 
           (in comparison with the other masses in the problem)
           lepton mass gives the full answer for the cross-section. The Drell-Yan form of the spectrum is restored 
           with such parameters, so the radiative corrections are negligible. But if the intermediate particle is rather light 
           than effect of the tau-lepton is not so crucial (Fig.\ref{s}d). Thus we have shown than one should not ignore 
           the "small" lepton mass in the set of cases.\\    
            
        We show the differential cross-section for the Dirac incoming particles in Fig.\ref{s}(e,f) for the comparison.

      {\bf 2})
          The cross-section averaged over both electron and positron spectra for the Dirac:
       \ba
        \sigma_D=\sigma_D^B(1+\frac{\alpha}{\pi}K_D(a))
       \label{sigma_D}
       \ea 
          and the Majorana incoming particles:
          \begin{equation}
	     \sigma_M=\left\{
               \begin{array}{lr}
                \frac{\alpha |g_L|^4}{2^7\pi^2 v a  M^2}
                            \Pi_M                             ,& \ \mu^2\ll 1 \\
             \sigma_M^B (1+\frac{\alpha}{\pi}K_M(a)) ,\ & a\gg 2,
                \end{array}
             \right. б
              \label{sigma_M}
           \end{equation}
         with
          $$
           \Pi_M = \left( 4a-1 
            -\frac{a((11-4a)a-6)}{2a-2}\ln\left[\frac{a-2}{a}\right]\right.\nn \\
            +\left.a^2\left(\frac{\pi^2}{6}-\ln^2\left[\frac{a}{2a-2}\right] 
          -2\mathcal{L}i_2 \left[\frac{a}{2a-2}\right] \right)
            \right).
          $$
      We note that K-factor is logarithmically divergent in the limit of \mbox{$a\to2$}, i.e.
          \ba
            K_i(a) = -\frac{1}{2}\ln^2|a-2|+O(\ln|a-2|).
          \ea
    Remember that $a=1+\frac{M_s^2}{M^2}$, where $M_S$ and $M$ are masses of the intermediate and incoming
    particles, respectively. And one can obtain that K-factor blows up and exceeds another terms when
         \ba
         \mu^2 \ln^2|a-2|\approx 1 \quad \mbox{in other words}\nn \\
         \Delta M= \frac{M_S^2 -M^2}{M^2}\approx e^{-\mu}
         \ea
    In this regime our result is not applicable. But physically it is almost not possible to reach it.\\

      {\bf 3}) Now we should  remember about full interaction Lagrangian in the eq.(\ref{majorana_lagrangian}). 
        Performing all previous calculations with respect to $g_L$ and $g_R$ terms one can get that
        $|g_L|^4\to(|g_L|^2+|g_R|^2)^2$ for the Born cross-sections:
       $$
          \sigma_M^B=\frac{(|g_L|^2+|g_R|^2)^2 m^2}{2^6\pi v a^2 M^4},\quad
          \sigma_D^B=\frac{(|g_L|^2+|g_R|^2)^2 }{2^6\pi v a^2 M^2},\quad
           $$
         with conservation of the renormalization group structures of equations (\ref{sigma_M}) and (\ref{sigma_D}).
       Because of $g_L^2g_R^2$ terms proportional to $m^2$ for the Majorana case, eq.(\ref{f0}) and its sequences 
      take the form:
         \ba
            F_0=\frac{\alpha(|g_L|^4+|g_R|^4)(1-x)}{2^4 \pi^2 v\tau_+^2\tau_-^2 M^2 }\left[ (1-x_+)^2+(1-x_-)^2\right].
                 \ea
         K-factor is slightly changed.\\
       {\bf 4}) 
	    Being integrated over the  lepton energy fractions all mass singularities disappears,
	    which  is in the agreement with the Kinoshita-Lee-Nauenberg theorem \cite{KNL}.

	    Above we had considered only t-channel Feynman amplitudes. Taking into account the
	    s-channel  diagrams with intermediate state of heavy vector neutral bosons
	    the additional terms of order $m^2/M^2$  for the case of Dirac particles will appear.
	    These terms will not change the results given above. For the case of Majorana particles
	    it results in some modification of the Born cross-section still proportional to the squared
	    lepton mass.
           
         
               \acknowledgements
	     We are
	    grateful to Yu.~M.~Bystritskiy for the independent check of the numerical results,
	    to Andrew Semenov for the attention to these problems and to Lars Bergstrom and Joakim Edsjo
            for valuable
	    criticism. One of us (R.A.) acknowledges support from the RFBR grant 08-02-00856-a. 

	    \section{Appendix}
            \subsection*{Properties of Majorana spinors}

	    We put here the necessary relations for the Majorana spinors \cite{CHUNG}:
	    \ba
	    \psi^c=C\bar{\psi}^T; C^+=C^{-1}; C^T=-C; C^{-1}\gamma^\mu C=-(\gamma^\mu)^T; \nn \\
	    u(k,s)=C\bar{v}^T(k,s);v(k,s)=C\bar{u}^T(k,s).
	    \ea
	    Bilinear combinations of Majorana spinors:
	    \ba
	    \sum_s u(p,s)\bar{u}(p,s)=\hat{p}+m; \nn \\
	    \sum_s v(p,s)\bar{v}(p,s)=\hat{p}-m.
	    \ea
	    \ba
	    \sum_s u(p,s)v^T(p,s)=-(\hat{p}+m)C; \nn \\
	    \sum_s v(p,s)u^T(p,s)=-(\hat{p}-m)C; \nn \\
	    \sum_s \bar{u}^T(p,s)\bar{v}(p,s)=C^+(\hat{p}-m); \nn \\
	    \sum_s \bar{v}^T(p,s)\bar{u}(p,s)=C^+(\hat{p}+m).
	    \ea
           \subsection*{K-factors} 
           
            Let us write down all mentioned  K-factors here:\\
            1) From calculation of the hard photon contribution we have:
            \ba
             \Phi_M(x_+,x_-)&=& \frac{4[a^2 -2ax+2(1-x_+)(1-x_-)][a(1+(1-x)^2)-4x(1-x)]}{\tau_+^2 \tau_-^2 (2x-a)},\\
             \Phi_D(x_+,x_-)&=& \frac{1}{\tau_+^2\tau_-^2}(-a^4 +2a^3(x+2) +8 a x(2+x(x-2))\nn\\
                &&8(2+x(x-2))(1-x_+)(1-x_-)-2a^2(4+x^3+x(x_++x_--2x_+x_-))),\\
            && x=2-x_+-x_-.\nn
            \ea
            2)From double differential distribution eq.(\ref{ddd}):
             \ba
	    K_i^h(x_+,x_-)=\frac{1}{2}[\delta(1-x_+)(1-x_-+\frac{1+x_-^2}{1-x_-}\ln \frac{x_-^2}{1-x_-})+\nn\\
	    \delta(1-x_-)(1-x_++\frac{1+x_+^2}{1-x_+}\ln \frac{x_+^2}{1-x_+})]+\Phi_i(x_+,x_-),
	    \ea
            3)From positron eq.(\ref{ds_dxplus}):
            \ba
	    K_i(x_+)&=&\frac{\pi^2}{3}-\frac{3}{4}+\frac{1}{2}[1-x_++\frac{1+x_+^2}{1-x_+}\ln \frac{x_+^2}{1-x_+}]+ \nn \\
	    &+&\int\limits_{1-x_+}^1\Phi_i(x_+,x_-)dx_-  
           \label{DefK}
	    \ea
              and photon (eq.(\ref{phod})) energy distributions
	    \ba
	    K^\gamma_i(x)&=&\frac{\pi^2}{3}-\frac{3}{4}+x+\frac{1+(1-x)^2}{x}\ln(1-x)+
	    \int\limits_{1-x}^1\Phi_i(2-x-x_-,x_-)dx_-. \label{DefKGamma}
	    \ea
        K-factors are small enough to be absent because they are proportional to the lepton  mass and do not have 
        any enhancement factors.

	    \end{document}